\documentclass[twocolumn,prd,longbibliography]{revtex4}
\usepackage{graphicx}
\usepackage{amssymb,amsfonts,mathrsfs,amsmath}
\newcommand{\tr}{\mbox{Tr}}

\newcommand{\CF}{{\cal F}}

\newcommand{\CZ}{{\cal Z}}

\newcommand{\be}{\begin{equation}}
\newcommand{\ee}{\end{equation}}
\newcommand{\bea}{\begin{eqnarray}}
\newcommand{\eea}{\end{eqnarray}}
\usepackage{graphicx}
\newcommand{\SC}{{\mathscr C}}

\newcommand{\SJ}{{\mathscr J}}
\newcommand{\SM}{{\mathscr M}}
\newcommand{\sigp}{{\Sigma^+}}
\newcommand{\sigm}{{\Sigma^-}}
\newcommand{\sigw}{{\Sigma^\wedge}}
\newcommand{\cl}{{\rm cl}}
\newcommand{\qu}{{\rm qu}}

\newcommand{\T}{{\rm T}}
\newcommand{\ie}{{\it i.e.}}
\newcommand{\eg}{{\it e.g.}}
\begin{document}
%%%%%%%%%%%%%%%%%%%%%%%%%%%%%%%%%%%%%%%%%%%%%%%%%%%%%%%%%%%%%%%%%%%%%%%%%%%%%%%
\title{String Perturbation Theory on the Schwinger-Keldysh Time Contour}
\author{Petr Ho\v{r}ava and Christopher J. Mogni}
\affiliation{\smallskip
Berkeley Center for Theoretical Physics and Department of Physics\\ 
University of California, Berkeley, California 94720-7300\\
and\\
Physics Division, Lawrence Berkeley National Laboratory\\ 
Berkeley, California 94720-8162}
\begin{abstract}
We perform the large-$N$ expansion in the Schwinger-Keldysh formulation of non-equilibrium quantum systems with matrix degrees of freedom, and study universal features of the anticipated dual string theory. We find a rich refinement of the topological genus expansion: In the original formulation, the future time instant where the forward and backward branches of the Schwinger-Keldysh time contour meet is associated with its own worldsheet genus expansion.  After the Keldysh rotation, the worldsheets decompose into a classical and quantum part.
\end{abstract}
\maketitle
%%%%%%%%%%%%%%%%%%%%%%%%%%%%%%%%%%%%%%%%%%%%%%%%%%%%%%%%%%%%%%%%%%%%%%%%%%%%%%%
\begin{center}
  \textbf{I. Introduction}
\end{center}
Non-equilibrium many-body systems are of central interest in remarkably many areas of physics, across a vast range of scales:  From the micoscopic scales of particle physics, to mesoscopic phenomena and condensed matter physics, to the cosmological scales of the cosmic microwave background and the large-scale structure of the Universe.  Moreover, the fluctuations governing the collective behavior in such systems may be either quantum or classical, thermal in nature.  

In the past few decades, the paradigm of string theory has proven to be a powerful generator of novel theoretical concepts which have found their way into remarkably many areas of physics and mathematics, not only to quantum gravity and particle phenomenology beyond the standard model, but also to condensed matter in holographic dualities and AdS/CFT correspondence \cite{zaanen,hartnoll}, or in helping with the topological classification of new topological states and phases of matter \cite{kth}.  One naturally wonders, can this useful influence of string theory be extended to non-equilibrium systems?

A direct attempt to formulate string theory far away from equilibrium faces a strong, historically rooted obstacle:  String theory originated \cite{birth} from the theory of the S-matrix, which is itself strongly based on the assumption of the static, stable, eternal relativistic vacuum. While a long list of brilliant results have been accumulated for strings out of equilibrium using traditional methods (in particular, in string cosmology), they have been achieved \textit{in spite of} this obstacle.  The purpose of this Letter is to propose steps towards eliminating this obstacle from first principles.  Having access to a broader formulation of string theory away from equilibrium would benefit a number of subdisciplines, from early-universe cosmology, the physics of black holes and the information puzzle, to less traditional areas for applied string theory, such as nonequilibrium mesoscopic physics.  A duality to string theory may provide access to a new weakly-coupled perturbative technique for otherwise strongly-correlated nonequilibrium systems, similar to what we witnessed in the context of AdS/CMT correspondence \cite{zaanen,hartnoll}.

Generally, in many-body physics such an assumption is far from necessary.  The more general formulation, which could simply be called ``quantum mechanics without simplifying assumptions'' about the vacuum, is known as the Schwinger-Keldysh (SK) formalism \cite{schwinger,keldysh} (see \cite{neq} for a comprehensive list of reviews).  The system is evolved forward and then backward, along a doubled time contour called the Schwinger-Keldysh (SK) time contour.  Equivalently, this doubling can be viewed as a doubling of fields on the single-valued time $t$.  This formalism has been the leading go-to technique for handling non-equilibrium many-body systems in condensed matter and a broad range of related areas for many decades.   It also plays an increasingly important role in gravity and cosmology, which goes back to the early pioneering and insightful work by H\'aj\'\i \v cek \cite{hajicek}, and later by Jordan \cite{jordan}.  In this century, the importance of the SK ``in-in'' formalism for inflationary cosmology has been particularly stressed by Weinberg \cite{weinberg1,weinberg2,weinberg3} (see also \cite{baumann}).  

\begin{figure}[b!]
\centering
\includegraphics[width=1.8in]{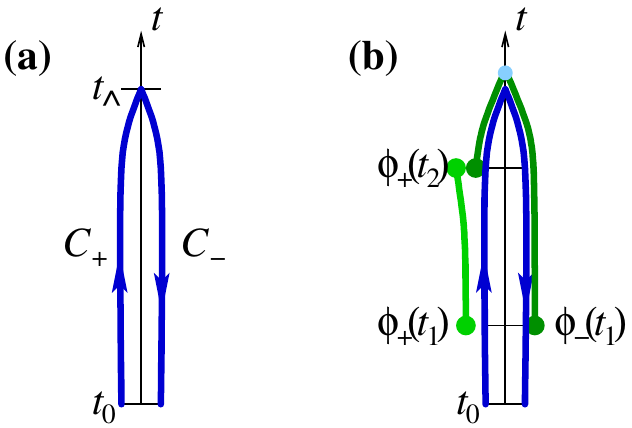}
\caption{{\bf (a):} The Schwinger-Keldysh contour $\SC=C_+\cup C_-$. {\bf (b):} Worldlines of particles corresponding to two of the four $G_{\pm\pm}$ propagators: $\langle\phi_+(t_1)\phi_+(t_2)\rangle_0$ and $\langle\phi_-(t_1)\phi_+(t_2)\rangle_0$.} 
\label{ff0}
\end{figure}

Understanding how string theory relates to the SK formalism is an important step towards developing non-equilibrium string theory.  Some work on SK formalism and strings already exists in the literature, primarily from the spacetime point of view \cite{kostas1,kostas2,jandb,haehl1,haehl2,sonh}.  Here we follow a different strategy:  In equilibrium, the structure of the large-$N$ expansion in theories with interacting matrix degrees of freedom predicts a string coupling expansion, as a sum over connected worldsheet topologies $\Sigma$ of increasing genus,  
\be
\CZ=\sum_{h=0}^\infty\left(\frac{1}{N}\right)^{2h-2}\CF_h(\lambda,\ldots).
\label{eenws}
\ee
Here $g_s\equiv 1/N$ plays the role of the string coupling constant, with the power of $N$ given by the Euler number $\chi(\Sigma)=2-2h$.  The 't~Hooft coupling $\lambda$ is a worldsheet coupling analogous to $\alpha'$ of critical strings, with ``$\ldots$'' suggesting there might be more than one such worldsheet coupling.  The importance of this duality, first developed by 't~Hooft \cite{th,th2}, was further advocated over the years by Polyakov (see, \eg , \cite{polyakov}), and others.  It turned out that at least in special circumstances, when combined with additional spacetime symmetries, this relationship reveals a lot about the dual string theory, eventually leading to such milestones as AdS/CFT correspondence \cite{juan}.  Can this relationship be extended away from equilibrium, and if so, what does it reveal about the perturbative expansion of the dual string theory?

The advantage of asking this question first on the large-$N$ side is that we understand conceptually quite well how to take that system out of equilibrium:  Simply apply the SK formalism.  On the string dual side, much less is known about non-equilibrium, and we can hope to learn something new by taking the correspondence seriously.  

We started this exploration of the large-$N$ expansion in non-equilibrium in our recent paper \cite{neq}, using the original ``forward-backward'' (or ``$\pm$'' for short) version of the SK formalism.  The genus expansion of equilibrium string theory is refined, into a sum over triple decompositions of the worldsheet topologies.  We present our main results below in Section~II, without proofs; the detailed arguments can be found in \cite{neq}. 

In many physical applications, the $\pm$ version of the SK formalism is found to be a little cumbersome \cite{vilkovisky,rammer}, and it is convenient to perform a redefinition of fields known as the Keldysh rotation.  In Section~III, we extend our analysis from \cite{neq} to the Keldysh-rotated formalism.  The large-$N$ expansion then predicts an intriguing new structure of the surfaces: Each $\Sigma$ decomposes into a classical and quantum part.  We again leave out all proofs; more details can be found in our forthcoming paper \cite{keq}.  In Section~IV, we specialize our attention to the class of models in which the Feynman ribbon diagrams give a geometric discretization of the metric properties of $\Sigma$, as \eg\ in the old matrix models for noncritical strings.  Our refined string perturbation theory then exhibits some additional special features, not necessarily shared by all string theory duals in the more general case. 
\smallskip
\begin{center}
  \textbf{II. Non-Equilibrium String Perturbation Theory\\ from Large $N$}
\end{center}
How can worldsheets be mapped to the target spacetime which incorporates the SK time contour?  Fig.~\ref{ff0}(b) gives some first intuition:  When particle worldlines are mapped to the SK contour, the worldlines corresponding to the propagators whose ends are at the two branches of the time contour must cross the point where the two branches meet, and that can be represented graphically by a cut across the propagator.  So perhaps when worldsheets map to the SK time contour, they should exhibit similar cuts.

In the large-$N$ duality, worldsheets are made from ribbon diagrams.  Consider the quantum many-body system of Hermitian matrices $M^a{}_b(t,\ldots)$, in the adjoint of the symmetry group $SU(N)$.   The ``$\ldots$'' here refer to all dependence of $M_\pm$ on spatial coordinates or additional quantum numbers, which we keep implicit, indicating only the dependence on time.  Thus, our results will be universal, regardless of whether $M$ are spacetime fields (such as Yang-Mills gauge fields), relativistic or not, or just matrices in a simple quantum mechanics: Our conclusions will be the same.

We assume that in equilibrium, the action is of the single-trace form,
\be
S(M)=\frac{1}{g^2}\int dt\,\tr\left(M^2+M^3+M^4+\ldots\right),
\ee
and we study it in the $N\to\infty$ limit with the 't~Hooft coupling $\lambda=g^2N$ held fixed.  Famously, this expansion leads to a dual perturbative expansion into string worldsheet topologies, Eqn.~(\ref{eenws}).  

Now we extend this duality away from equilibrium, considering the system of $M$ in the SK formalism.  First, we have the usual doubling of fields to $M_+(t,\ldots)$ and $M_-(t,\ldots)$, representing the values of $M$ at the forward and backward branch of the SK contour.  In this ``forward-backward'' formalism, the action that reproduces the diagrammatic rules away from equilibrium can be succinctly written as 
\be
S_{\rm SK}=S(M_+)-S(M_-),
\ee
although one needs to exercise some care about appropriate conditions where the two parts $C_\pm$ of the time contour meet \cite{kamenev}.  

The Feynman rules contain four types of propagators, 
\bea
\vcenter{\hbox{\includegraphics[width=.65in]{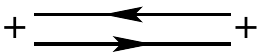}}} &=&\left\langle\T_\SC\left(M_+M_+\right)\right\rangle_0=(\lambda/N)\,G_{++},
\label{eeproppp}
\\
\vcenter{\hbox{\includegraphics[width=.65in]{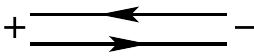}}} &=&\left\langle\T_\SC\left( M_+M_-\right)\right\rangle_0=(\lambda/N)\,G_{+-},
\label{eeproppm}
\\
\vcenter{\hbox{\includegraphics[width=.65in]{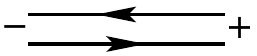}}} &=&\left\langle\T_\SC\left( M_-M_+\right)\right\rangle_0=(\lambda/N)\,G_{-+},
\label{eepropmp}
\\
\vcenter{\hbox{\includegraphics[width=.65in]{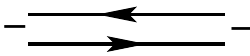}}} &=&\left\langle\T_\SC\left( M_-M_-\right)\right\rangle_0=(\lambda/N)\,G_{--},
\label{eepropmm}
\eea
with $\T_\SC$ denoting the time ordering along the contour $\SC$.  Each vertex gets labeled by a sign,
\bea
\vcenter{\hbox{\includegraphics[width=.3in]{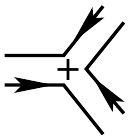}}} = -\frac{iN}{\lambda},&&\qquad
\vcenter{\hbox{\includegraphics[width=.3in]{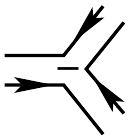}}} = \frac{iN}{\lambda},\\
\vcenter{\hbox{\includegraphics[width=.35in]{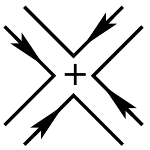}}} = -\frac{iN}{\lambda},&&\qquad
\vcenter{\hbox{\includegraphics[width=.35in]{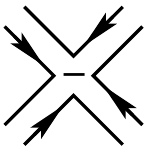}}} = \frac{iN}{\lambda}.
\label{eevert--}
\eea
Note that we have also suppressed the $SU(N)$ indices $a,b,\ldots$.  Unlike all the other indices and spatial dependences that we keep implicit and whose role is largely decorative, the $SU(N)$ indices are of course crucial to our arguments, but they are well-represented in the usual way in the graphical form:  They are carried by the edges of the ribbon diagrams.  

Thus, compared to equilibrium, the novelty of the ribbon Feynman diagrams in non-equilibrium is in the labeling of vertices as $+$ or $-$.  This extra structure in turn induces additional topological features on the associated worldsheet surfaces.  Certainly, there will be portions $\sigp$ and $\sigm$ of $\Sigma$ such that their vertices and propagators reside respectively either solely on $C_+$  or solely on $C_-$.  These two surfaces should then be joined to form $\Sigma$.  The question is, how are they joined?  One can begin by indicating each propagator which straddles the two branches of the SK contour (\ie , $G_{+-}$ and $G_{-+}$), by putting a cut across it, to indicate that it crosses from the $+$ region to the $-$ region: 
\be
\vcenter{\hbox{\includegraphics[width=.65in]{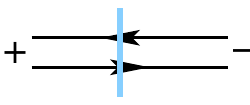}}}\ .
\nonumber
\ee
Then one can try to extend these propagator cuts into consistent cuts of the entire worldsheets.  

A detailed analysis reveals that such cuts cannot be uniquely and usefully extended \cite{neq}.  More precisely, when extended in a unique way across $\Sigma$, they become complicated graphs on $\Sigma$, not just a collection of a few $S^1$ boundaries between $\sigp$ and $\sigm$.  To resolve this issue, one needs to widen the cut into a portion of a two-dimensional surface with a boundary, 
\be
\vcenter{\hbox{\includegraphics[width=.65in]{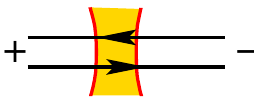}}}\ .
\nonumber\ee
It is these widened cuts that connect smoothly and uniquely across all plaquettes, and lead to a decomposition of $\Sigma$ into {\it three\/} topologically meaningful parts characterized by natural topological invariants.  In this three-fold decomposition, the region of $\Sigma$ that corresponds to the instant $t_\wedge$ in time where the two branches of the SK time contour meet is not just a collection of $S^1$ boundaries between $\sigp$ and $\sigm$:  It is a two-dimensional surface $\sigw$, with its own nontrivial topology, and its own genus expansion!

Combinatorially, we define this triple topological decomposition of $\Sigma$ as follows.  All the $+$ vertices, $G_{++}$ propagators, and all the plaquettes whose all adjacent propagators are $G_{++}$ define $\sigp$; analogously for $\sigm$.  Finally, all $G_{+-}$ and $G_{-+}$ propagators and all the plaquettes with at least one such $G_{\pm\mp}$ propagator define the wedge region $\sigw$.  A careful analysis shows \cite{neq} that all topologies of the three regions do indeed appear from consistent ribbon diagrams, as long as they respect that their union $\Sigma$ is connected.

Thus, we conclude that in the non-equilibrium case, the genus expansion (\ref{eenws}) of string perturbation theory is refined into a sum over triple decompositions of worldsheets,
\begin{figure}
\centering
\includegraphics[width=1.8in]{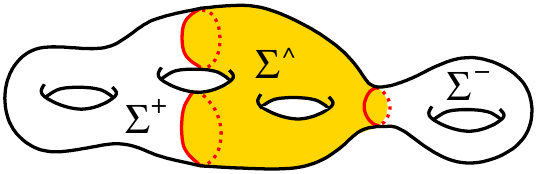}
\caption{A typical surface $\Sigma$ and its triple decomposition into the forward region $\sigp$, the backward region $\sigm$, and the wedge region $\sigw$ connecting them.} 
\label{ffii}
\end{figure}
\be
\CZ=\sum_{h=0}^\infty\left(\frac{1}{N}\right)^{2h-2}\!\!\!\!\!\!\!\sum_{\substack{{\rm triple\ decompositions}\\
\chi(\sigp\cup\sigm\cup\sigw)=2-2h}}
\!\!\!\!\!\CF_{\sigp,\sigm,\sigw}(\lambda,\ldots).\nonumber
\label{eenwsii}
\ee
A typical worldsheet that contributes to this sum is shown in Fig.~\ref{ffii}.
\smallskip
\begin{center}
  \textbf{III. Keldysh Rotation and\\ Non-Equilibrium String Perturbation Theory}
\end{center}
The Keldysh rotation is defined by a simple but useful linear change of variables, from the ``forward'' and ``backward'' fields $M_\pm$ to their ``classical'' and ``quantum'' counterparts,
\be
M_\cl=\frac{1}{2}(M_++M_-),\quad
M_\qu=\frac{1}{2}(M_+-M_-).
\label{eekrot}
\ee
To avoid clutter, we will simply call them $M_\cl\equiv M$ and $M_\qu\equiv\SM$.

In these new variables, the quadratic part of the SK action simplifies, only three propagators are nonzero:
\bea
\left\langle\SM\,M\right\rangle_0&=&G_A,\qquad \left\langle M\,\SM\right\rangle_0=G_R,\nonumber\\
\left\langle M\,M\right\rangle_0&=&G_K,\qquad \left\langle\SM\,\SM\right\rangle_0\equiv 0.\nonumber
\eea
Moreover, the information about the dynamics and about the state is now nicely separated: The advanced and retarded propagators $G_A$ and $G_R$ describe the dynamics, while the entire information about the state is encoded in the ``Keldysh propagator'' $G_K$.  This makes the Keldysh rotation both computationally more efficient, and physically more intuitive \cite{rammer,svl}.

We similarly rotate the sources,
\be
J_\cl\equiv J=\frac{1}{2}(J_++J_-),\quad
J_\qu\equiv\SJ=\frac{1}{2}(J_+-J_-).
\ee

To simplify the contruction of Feynman diagrams, we find it useful to introduce a ``signpost notation'' \cite{keq}: Each vertex is equipped with a collection of arrows (\ie , a ``signpost''), one pointing into each quantum end of the vertex.  The vertices are:
\bea
\vcenter{\hbox{\includegraphics[width=.35in]{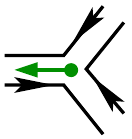}}}\ ,&&\qquad
\vcenter{\hbox{\includegraphics[width=.35in]{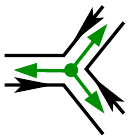}}}\ ,\nonumber\\
\vcenter{\hbox{\includegraphics[width=.4in]{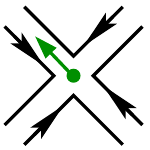}}}\ ,&&\qquad
\vcenter{\hbox{\includegraphics[width=.4in]{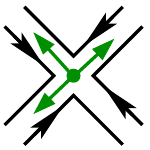}}}\ .\nonumber
\eea
Note that in full generality, $S_{\rm SK}$ always gives only vertices with an odd number of $\SM$ ends (and hence an odd number of arrows in their signpost).  The Feynman rules are such that propagators must match the signposts: Each $\langle\SM\,M\rangle_0$ and $\langle M\,\SM\rangle_0$ propagator has an arrow pointing into it at its quantum end.  Furthermore, if any diagram contains a closed path formed by following the arrows at each vertex, it is zero identically \cite{keq}.

The Keldysh rotation (\ref{eekrot}) is a simple transformation of variables, but the two string dual pictures are not expected to be related to each other in any simple way \cite{keq}:  The relation between the worldsheets $\Sigma$ before and after the Keldysh rotation would require a prohibitively complicated resummation of many ribbon diagrams.  Thus, we find two distinct dual string expansions, each with its unique features.

In the Keldysh-rotated formalism for the large-$N$ expansion, we find that the sum over surface topologies is again refined, but in a different way \cite{keq}:  Each $\Sigma$ is decomposed into its {\it classical part\/} $\Sigma^\cl$ and {\it quantum part\/} $\Sigma^\qu$.  Combinatorially, these parts are defined as follows: All $G_A$ and $G_R$ propagators, all internal vertices, and all the plaquettes without adjacent $G_K$ propagators define the classical part $\Sigma^\cl$.  All the $G_K$ propagators and all the plaquettes with at least one adjacent $G_K$ propagator comprise the quantum part $\Sigma^\qu$.  Note that the nice separation of the dynamics from the many-body state of the large-$N$ system extends to the string side:  The information about the state resides in $G_K$, and is now entirely carried by $\Sigma^\qu$.  In contrast, $\Sigma^\cl$ contains all the elements that only know about the dynamics.

Further analysis shows \cite{keq} that all possible topologies of $\Sigma^\cl$ and $\Sigma^\qu$ emerge from consistent ribbon diagrams, as long as they give a connected $\Sigma$.  We conclude that in the Keldysh-rotated description, the sum over topologies in non-equilibrium string perturbation theory is refined to a sum over the worldsheet decompositions into their classical and quantum parts, 
\begin{figure}[t!]
\centering
\includegraphics[width=1.8in]{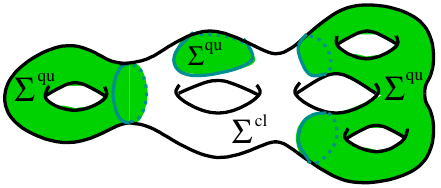}
\caption{A typical surface $\Sigma$ and its decomposition into the classical and quantum part $\Sigma^\cl$ and $\Sigma^\qu$.} 
\label{ffiii}
\end{figure}
\be
\CZ(J,\SJ)=\sum_{h=0}^\infty\left(\frac{1}{N}\right)^{2h-2}\!\!\!\!\!\!\!\sum_{\substack{{\rm double\ decompositions}\\
\chi(\Sigma^\cl)+\chi(\Sigma^\qu)=2-2h}}
\!\!\!\!\!\!\!\CF_{\Sigma^\cl,\Sigma^\qu}(\lambda;J,\SJ).\nonumber
\label{eenwsiii}
\ee
We have added the dependence on the sources, because all the individual vacuum diagrams can be shown to be identically zero \cite{keq}.  This equation then holds for all the correlation functions of $M$ and $\SM$.  A typical example of a surface contributing to the sum is shown in Fig.~\ref{ffiii}.
\begin{center}
  \textbf{IV. Random Triangulations}
\end{center}
The specific coefficients $\CF$ in our triple- and double-decomposition sums encode all the worldsheet dynamics, and are therefore generally out of reach for our universal analysis of the string loop expansion.

For special subclasses of models, however, we can obtain further insights, without losing the universal nature of our conclusions.  Consider the subclass consisting of those large-$N$ systems, for which the ribbon diagrams serve as a tool to discretize the path-integral sum over worldsheet geometries and to regulate the Einstein-Hilbert action with a cosmological constant (coupled perhaps to simple types of worldsheet matter).  A typical example is given by the ``old-fashioned'' matrix models of nonperturbative noncritical string theories in $d\leq 1+1$ spacetime dimensions (see \cite{yu,igor,gm} for reviews).  Even though much has been learned about the matrix models from the more modern perspective of D-branes and tachyon condensation \cite{newhat,mcgv} (see also \cite{yu}), the original reasoning that uses ribbon diagrams to define the worldsheet path integral as a sum over random triangulations is still perfectly valid.  In this construction, each Feynman ribbon diagram is viewed as a dual to a triangulation (or more generally, a simplicial decomposition) of $\Sigma$.  One unit $A$ of discretized area is assigned to each {\it vertex\/} of the large-$N$ ribbon diagram, \ie , to each {\it plaquette\/} of the dual diagram.   The regulated worldsheet path integral is then given by the sum over all Feynman diagrams, and its continuum limit is described by a cetain double scaling limit of the large-$N$ theory.

With this reminder, we are ready to identify additional universal features of the topological genus expansion of non-equilibrium string theory, in this special class of models.  Recall how we assigned the various combinatorial elements of the ribbon diagrams in the $\pm$ formalism to the three parts of the triple decomposition of $\Sigma$:  All vertices were naturally assigned to $\sigp$ and $\sigm$, with the wedge region $\sigw$ consisting only of certain propagators and plaquettes.  When one unit $A$ of the geometric area is ascribed to each vertex, this implies that -- despite being topologically two-dimensional -- the wedge region $\Sigma^\wedge$ of the worldsheet has zero regulated area!  It is likely that this feature should persist also in the continuum limit.

The same is true in the Keldysh-rotated formalism about the quantum part $\Sigma^\qu$ of the worldsheet:  Only certain plaquettes and propagators were assigned to $\Sigma^\qu$ in our combinatorial construction, with all vertices belonging to $\Sigma^\cl$.  Thus, $\Sigma^\qu$ carries zero regulated total area.  We reach a fairly universal conclusion, valid for the special broad class of random triangulations models: In the dual string theory, $\sigw$ and $\Sigma^\qu$ are topologically two-dimensional and carry their own genus expansion, yet geometrically they effectively appear to be at most one-dimensional objects (unless they overcome this tendency by acquiring large anomalous dimensions in the continuum limit).
\begin{center}
  \textbf{V. Conclusions}
\end{center}
In two distinct formulations of the SK formalism for the large-$N$ expansion of non-equilibrium systems with matrix degrees of freedom, we found that each leads to a distinct refinement of the sum over worldsheet topologies in string perturbation theory.  New topological invariants are now available (such as the Euler numbers of $\sigp$, $\sigm$ and $\sigw$ in the original $\pm$ formalism, or of $\Sigma^\cl$ and $\Sigma^\qu$ in the Keldysh-rotated formalism), and it is natural to expect that they can be weighted by different values of the string coupling $g_+$, $g_-$ and $g_\wedge$ (or $g_\cl$ and $g_\qu$).  Indeed, we envision a situation much like in equilibrium, where in some solutions, the string coupling $g_s$ can be spacetime dependent.  In the SK context, a time-dependent $g_s$ could lead to different effective values of the string coupling in the corresponding regions of the SK time contour.  

In order to gain further insights into such open questions, it is important to look at specific dynamical examples of string theory, where some aspects of the worldsheet dynamics can be controlled from first principles, and the predicted structure of non-equilibrium string perturbation theory can be directly verified.  A particularly natural candidate would be the maximally supersymmetric AdS/CFT correspondence:  Revisiting its origin from coincident D-brane systems, now taken out of equilibrium, could provide a bridge between the relatively well-understood spacetime picture \cite{kostas1,kostas2,jandb,haehl1,haehl2,sonh} and the sought-for worldsheet description of the Schwinger-Keldysh formalism for nonequilibrium systems.

%%%%%%%%%%%%%%%%%%%%%%%%%%%%%%%%%%%%%%%%%%%%%%%%%%%%%%%%%%%%%%%%%%%%%%%%%%%%%%%
{\bf Acknowledgements:}
This work has been supported by NSF grants PHY-1820912 and PHY-1521446.
%%%%%%%%%%%%%%%%%%%%%%%%%%%%%%%%%%%%%%%%%%%%%%%%%%%%%%%%%%%%%%%%%%%%%%%%%%%%%%%
\bibliography{neq}

\begin{thebibliography}{33}
\expandafter\ifx\csname natexlab\endcsname\relax\def\natexlab#1{#1}\fi
\expandafter\ifx\csname bibnamefont\endcsname\relax
  \def\bibnamefont#1{#1}\fi
\expandafter\ifx\csname bibfnamefont\endcsname\relax
  \def\bibfnamefont#1{#1}\fi
\expandafter\ifx\csname citenamefont\endcsname\relax
  \def\citenamefont#1{#1}\fi
\expandafter\ifx\csname url\endcsname\relax
  \def\url#1{\texttt{#1}}\fi
\expandafter\ifx\csname urlprefix\endcsname\relax\def\urlprefix{URL }\fi
\providecommand{\bibinfo}[2]{#2}
\providecommand{\eprint}[2][]{\url{#2}}

\bibitem[{\citenamefont{Zaanen et~al.}(2015)\citenamefont{Zaanen, Sun, Liu, and
  Schalm}}]{zaanen}
\bibinfo{author}{\bibfnamefont{J.}~\bibnamefont{Zaanen}},
  \bibinfo{author}{\bibfnamefont{Y.-W.} \bibnamefont{Sun}},
  \bibinfo{author}{\bibfnamefont{Y.}~\bibnamefont{Liu}}, \bibnamefont{and}
  \bibinfo{author}{\bibfnamefont{K.}~\bibnamefont{Schalm}},
  \emph{\bibinfo{title}{{Holographic Duality in Condensed Matter Physics}}}
  (\bibinfo{publisher}{Cambridge Univ. Press}, \bibinfo{year}{2015}).

\bibitem[{\citenamefont{Hartnoll et~al.}(2016)\citenamefont{Hartnoll, Lucas,
  and Sachdev}}]{hartnoll}
\bibinfo{author}{\bibfnamefont{S.~A.} \bibnamefont{Hartnoll}},
  \bibinfo{author}{\bibfnamefont{A.}~\bibnamefont{Lucas}}, \bibnamefont{and}
  \bibinfo{author}{\bibfnamefont{S.}~\bibnamefont{Sachdev}}
  (\bibinfo{year}{2016}), \eprint{arXiv:1612.07324}.

\bibitem[{\citenamefont{Ho\v{r}ava}(2005)}]{kth}
\bibinfo{author}{\bibfnamefont{P.}~\bibnamefont{Ho\v{r}ava}},
  \bibinfo{journal}{Phys. Rev. Lett.} \textbf{\bibinfo{volume}{95}},
  \bibinfo{pages}{016405} (\bibinfo{year}{2005}),
  \eprint{arXiv:hep-th/0503006}.

\bibitem[{\citenamefont{Cappelli et~al.}(2012)\citenamefont{Cappelli,
  Castellani, Colomo, and Di~Vecchia}}]{birth}
\bibinfo{editor}{\bibfnamefont{A.}~\bibnamefont{Cappelli}},
  \bibinfo{editor}{\bibfnamefont{E.}~\bibnamefont{Castellani}},
  \bibinfo{editor}{\bibfnamefont{F.}~\bibnamefont{Colomo}}, \bibnamefont{and}
  \bibinfo{editor}{\bibfnamefont{P.}~\bibnamefont{Di~Vecchia}}, eds.,
  \emph{\bibinfo{title}{{The Birth of String Theory}}}
  (\bibinfo{publisher}{Cambridge Univ. Press}, \bibinfo{address}{Cambridge,
  UK}, \bibinfo{year}{2012}).

\bibitem[{\citenamefont{Schwinger}(1961)}]{schwinger}
\bibinfo{author}{\bibfnamefont{J.~S.} \bibnamefont{Schwinger}},
  \bibinfo{journal}{J. Math. Phys.} \textbf{\bibinfo{volume}{2}},
  \bibinfo{pages}{407} (\bibinfo{year}{1961}).

\bibitem[{\citenamefont{Keldysh}(1964)}]{keldysh}
\bibinfo{author}{\bibfnamefont{L.}~\bibnamefont{Keldysh}},
  \bibinfo{journal}{Zh. Eksp. Teor. Fiz.} \textbf{\bibinfo{volume}{47}},
  \bibinfo{pages}{1515} (\bibinfo{year}{1964}).

\bibitem[{\citenamefont{Ho\v{r}ava and Mogni}(2020{\natexlab{a}})}]{neq}
\bibinfo{author}{\bibfnamefont{P.}~\bibnamefont{Ho\v{r}ava}} \bibnamefont{and}
  \bibinfo{author}{\bibfnamefont{C.~J.} \bibnamefont{Mogni}}
  (\bibinfo{year}{2020}{\natexlab{a}}), \eprint{arXiv:2008.11685}.

\bibitem[{\citenamefont{H\'{a}j\'{\i}\v{c}ek}(1979)}]{hajicek}
\bibinfo{author}{\bibfnamefont{P.}~\bibnamefont{H\'{a}j\'{\i}\v{c}ek}}, in
  \emph{\bibinfo{booktitle}{{The Second Marcel Grossmann Meeting on the Recent
  Developments of General Relativity}}} (\bibinfo{year}{1979}), p.
  \bibinfo{pages}{483}.

\bibitem[{\citenamefont{Jordan}(1986)}]{jordan}
\bibinfo{author}{\bibfnamefont{R.}~\bibnamefont{Jordan}},
  \bibinfo{journal}{Phys. Rev. D} \textbf{\bibinfo{volume}{33}},
  \bibinfo{pages}{444} (\bibinfo{year}{1986}).

\bibitem[{\citenamefont{Weinberg}(2005)}]{weinberg1}
\bibinfo{author}{\bibfnamefont{S.}~\bibnamefont{Weinberg}},
  \bibinfo{journal}{Phys. Rev. D} \textbf{\bibinfo{volume}{72}},
  \bibinfo{pages}{043514} (\bibinfo{year}{2005}),
  \eprint{arXiv:hep-th/0506236}.

\bibitem[{\citenamefont{Weinberg}(2006)}]{weinberg2}
\bibinfo{author}{\bibfnamefont{S.}~\bibnamefont{Weinberg}},
  \bibinfo{journal}{Phys. Rev. D} \textbf{\bibinfo{volume}{74}},
  \bibinfo{pages}{023508} (\bibinfo{year}{2006}),
  \eprint{arXiv:hep-th/0605244}.

\bibitem[{\citenamefont{Weinberg}(2008)}]{weinberg3}
\bibinfo{author}{\bibfnamefont{S.}~\bibnamefont{Weinberg}},
  \bibinfo{journal}{Phys. Rev. D} \textbf{\bibinfo{volume}{77}},
  \bibinfo{pages}{123541} (\bibinfo{year}{2008}), \eprint{arXiv:0804.4291}.

\bibitem[{\citenamefont{Baumann and McAllister}(2015)}]{baumann}
\bibinfo{author}{\bibfnamefont{D.}~\bibnamefont{Baumann}} \bibnamefont{and}
  \bibinfo{author}{\bibfnamefont{L.}~\bibnamefont{McAllister}},
  \emph{\bibinfo{title}{{Inflation and String Theory}}}
  (\bibinfo{publisher}{Cambridge University Press}, \bibinfo{year}{2015}),
  \eprint{arXiv:1404.2601}.

\bibitem[{\citenamefont{Skenderis and van Rees}(2008)}]{kostas1}
\bibinfo{author}{\bibfnamefont{K.}~\bibnamefont{Skenderis}} \bibnamefont{and}
  \bibinfo{author}{\bibfnamefont{B.~C.} \bibnamefont{van Rees}},
  \bibinfo{journal}{Phys. Rev. Lett.} \textbf{\bibinfo{volume}{101}},
  \bibinfo{pages}{081601} (\bibinfo{year}{2008}), \eprint{arXiv:0805.0150}.

\bibitem[{\citenamefont{Skenderis and van Rees}(2009)}]{kostas2}
\bibinfo{author}{\bibfnamefont{K.}~\bibnamefont{Skenderis}} \bibnamefont{and}
  \bibinfo{author}{\bibfnamefont{B.~C.} \bibnamefont{van Rees}},
  \bibinfo{journal}{JHEP} \textbf{\bibinfo{volume}{05}}, \bibinfo{pages}{085}
  (\bibinfo{year}{2009}), \eprint{arXiv:0812.2909}.

\bibitem[{\citenamefont{de~Boer et~al.}(2019)\citenamefont{de~Boer, Heller, and
  Pinzani-Fokeeva}}]{jandb}
\bibinfo{author}{\bibfnamefont{J.}~\bibnamefont{de~Boer}},
  \bibinfo{author}{\bibfnamefont{M.~P.} \bibnamefont{Heller}},
  \bibnamefont{and}
  \bibinfo{author}{\bibfnamefont{N.}~\bibnamefont{Pinzani-Fokeeva}},
  \bibinfo{journal}{JHEP} \textbf{\bibinfo{volume}{05}}, \bibinfo{pages}{188}
  (\bibinfo{year}{2019}), \eprint{arXiv:1812.06093}.

\bibitem[{\citenamefont{Haehl et~al.}(2017{\natexlab{a}})\citenamefont{Haehl,
  Loganayagam, and Rangamani}}]{haehl1}
\bibinfo{author}{\bibfnamefont{F.~M.} \bibnamefont{Haehl}},
  \bibinfo{author}{\bibfnamefont{R.}~\bibnamefont{Loganayagam}},
  \bibnamefont{and}
  \bibinfo{author}{\bibfnamefont{M.}~\bibnamefont{Rangamani}},
  \bibinfo{journal}{JHEP} \textbf{\bibinfo{volume}{06}}, \bibinfo{pages}{069}
  (\bibinfo{year}{2017}{\natexlab{a}}), \eprint{arXiv:1610.01940}.

\bibitem[{\citenamefont{Haehl et~al.}(2017{\natexlab{b}})\citenamefont{Haehl,
  Loganayagam, and Rangamani}}]{haehl2}
\bibinfo{author}{\bibfnamefont{F.~M.} \bibnamefont{Haehl}},
  \bibinfo{author}{\bibfnamefont{R.}~\bibnamefont{Loganayagam}},
  \bibnamefont{and}
  \bibinfo{author}{\bibfnamefont{M.}~\bibnamefont{Rangamani}},
  \bibinfo{journal}{JHEP} \textbf{\bibinfo{volume}{06}}, \bibinfo{pages}{070}
  (\bibinfo{year}{2017}{\natexlab{b}}), \eprint{arXiv:1610.01941}.

\bibitem[{\citenamefont{Herzog and Son}(2003)}]{sonh}
\bibinfo{author}{\bibfnamefont{C.~P.} \bibnamefont{Herzog}} \bibnamefont{and}
  \bibinfo{author}{\bibfnamefont{D.~T.} \bibnamefont{Son}},
  \bibinfo{journal}{JHEP} \textbf{\bibinfo{volume}{03}}, \bibinfo{pages}{046}
  (\bibinfo{year}{2003}), \eprint{arXiv:hep-th/0212072}.

\bibitem[{\citenamefont{'t~Hooft}(1974{\natexlab{a}})}]{th}
\bibinfo{author}{\bibfnamefont{G.}~\bibnamefont{'t~Hooft}},
  \bibinfo{journal}{Nucl. Phys. B} \textbf{\bibinfo{volume}{72}},
  \bibinfo{pages}{461} (\bibinfo{year}{1974}{\natexlab{a}}).

\bibitem[{\citenamefont{'t~Hooft}(1974{\natexlab{b}})}]{th2}
\bibinfo{author}{\bibfnamefont{G.}~\bibnamefont{'t~Hooft}},
  \bibinfo{journal}{Nucl. Phys. B} \textbf{\bibinfo{volume}{75}},
  \bibinfo{pages}{461} (\bibinfo{year}{1974}{\natexlab{b}}).

\bibitem[{\citenamefont{Polyakov}(1993)}]{polyakov}
\bibinfo{author}{\bibfnamefont{A.~M.} \bibnamefont{Polyakov}}, in
  \emph{\bibinfo{booktitle}{{Les Houches Summer School on Gravitation and
  Quantizations, Session 57}}} (\bibinfo{year}{1993}), p. \bibinfo{pages}{783},
  \eprint{hep-th/9304146}.

\bibitem[{\citenamefont{Maldacena}(1998)}]{juan}
\bibinfo{author}{\bibfnamefont{J.~M.} \bibnamefont{Maldacena}},
  \bibinfo{journal}{Adv. Theor. Math. Phys.} \textbf{\bibinfo{volume}{2}},
  \bibinfo{pages}{231} (\bibinfo{year}{1998}), \eprint{arXiv:hep-th/9711200}.

\bibitem[{\citenamefont{Vilkovisky}(2008)}]{vilkovisky}
\bibinfo{author}{\bibfnamefont{G.}~\bibnamefont{Vilkovisky}},
  \bibinfo{journal}{Lect. Notes Phys.} \textbf{\bibinfo{volume}{737}},
  \bibinfo{pages}{729} (\bibinfo{year}{2008}), \eprint{arXiv:0712.3379}.

\bibitem[{\citenamefont{Rammer}(2007)}]{rammer}
\bibinfo{author}{\bibfnamefont{J.}~\bibnamefont{Rammer}},
  \emph{\bibinfo{title}{{Quantum field theory of non-equilibrium states}}}
  (\bibinfo{publisher}{Cambridge University Press},
  \bibinfo{address}{Cambridge}, \bibinfo{year}{2007}).

\bibitem[{\citenamefont{Ho\v{r}ava and Mogni}(2020{\natexlab{b}})}]{keq}
\bibinfo{author}{\bibfnamefont{P.}~\bibnamefont{Ho\v{r}ava}} \bibnamefont{and}
  \bibinfo{author}{\bibfnamefont{C.~J.} \bibnamefont{Mogni}}
  (\bibinfo{year}{2020}{\natexlab{b}}), \eprint{arXiv:2010.10671}.

\bibitem[{\citenamefont{Kamenev}(2011)}]{kamenev}
\bibinfo{author}{\bibfnamefont{A.}~\bibnamefont{Kamenev}},
  \emph{\bibinfo{title}{{Field Theory of Non-Equilibrium Systems}}}
  (\bibinfo{publisher}{Cambridge University Press},
  \bibinfo{address}{Cambridge}, \bibinfo{year}{2011}).

\bibitem[{\citenamefont{Stefanucci and van Leeuwen}(2013)}]{svl}
\bibinfo{author}{\bibfnamefont{G.}~\bibnamefont{Stefanucci}} \bibnamefont{and}
  \bibinfo{author}{\bibfnamefont{R.}~\bibnamefont{van Leeuwen}},
  \emph{\bibinfo{title}{{Nonequilibrium Many-Body Theory of Quantum Systems}}}
  (\bibinfo{publisher}{Cambridge University Press},
  \bibinfo{address}{Cambridge}, \bibinfo{year}{2013}).

\bibitem[{\citenamefont{Nakayama}(2004)}]{yu}
\bibinfo{author}{\bibfnamefont{Y.}~\bibnamefont{Nakayama}}
  (\bibinfo{year}{2004}), \eprint{arXiv:hep-th/0402009}.

\bibitem[{\citenamefont{Klebanov}(1991)}]{igor}
\bibinfo{author}{\bibfnamefont{I.~R.} \bibnamefont{Klebanov}}, in
  \emph{\bibinfo{booktitle}{{Spring School on String Theory and Quantum
  Gravity}}} (\bibinfo{year}{1991}), p.~\bibinfo{pages}{30},
  \eprint{arXiv:hep-th/9108019}.

\bibitem[{\citenamefont{Ginsparg and Moore}(1993)}]{gm}
\bibinfo{author}{\bibfnamefont{P.~H.} \bibnamefont{Ginsparg}} \bibnamefont{and}
  \bibinfo{author}{\bibfnamefont{G.~W.} \bibnamefont{Moore}}, in
  \emph{\bibinfo{booktitle}{{Theoretical Advanced Study Institute (TASI 92):
  From Black Holes and Strings to Particles}}} (\bibinfo{year}{1993}), p.
  \bibinfo{pages}{277}, \eprint{arXiv:hep-th/9304011}.

\bibitem[{\citenamefont{Douglas et~al.}(2003)\citenamefont{Douglas, Klebanov,
  Kutasov, Maldacena, Martinec, and Seiberg}}]{newhat}
\bibinfo{author}{\bibfnamefont{M.~R.} \bibnamefont{Douglas}},
  \bibinfo{author}{\bibfnamefont{I.~R.} \bibnamefont{Klebanov}},
  \bibinfo{author}{\bibfnamefont{D.}~\bibnamefont{Kutasov}},
  \bibinfo{author}{\bibfnamefont{J.~M.} \bibnamefont{Maldacena}},
  \bibinfo{author}{\bibfnamefont{E.~J.} \bibnamefont{Martinec}},
  \bibnamefont{and} \bibinfo{author}{\bibfnamefont{N.}~\bibnamefont{Seiberg}}
  (\bibinfo{year}{2003}), \eprint{arXiv:hep-th/0307195}.

\bibitem[{\citenamefont{McGreevy and Verlinde}(2003)}]{mcgv}
\bibinfo{author}{\bibfnamefont{J.}~\bibnamefont{McGreevy}} \bibnamefont{and}
  \bibinfo{author}{\bibfnamefont{H.~L.} \bibnamefont{Verlinde}},
  \bibinfo{journal}{JHEP} \textbf{\bibinfo{volume}{12}}, \bibinfo{pages}{054}
  (\bibinfo{year}{2003}), \eprint{arXiv:hep-th/0304224}.

\end{thebibliography}
\end{document}